\documentstyle[twoside,fleqn,espcrc2]{article}

\input epsf

\newcommand{\beq}{\begin{equation}}
\newcommand{\eeq}{\end{equation}}
\newcommand{\bea}{\begin{eqnarray}}
\newcommand{\eea}{\end{eqnarray}}

\newcommand{\T}{\textstyle}

\newcommand{\tr}{{\rm tr}}

\newcommand{\V}{{\cal V}}
\newcommand{\vev}[1]{\Big\langle #1 \Big\rangle}

\newcommand{\AmS}{{\protect\the\textfont2
  A\kern-.1667em\lower.5ex\hbox{M}\kern-.125emS}}

\title{Vortex structure of the vacuum and confinement}

\author{T.G. Kov\'acs\address{Department of Physics, 
             Instituut-Lorentz for Theoretical Physics,\\ 
             P.O.Box 9506, 2300 RA, Leiden, The Netherlands}%
        \thanks{Present address: DESY-Zeuthen, D-15738 Zeuthen, Germany. 
         Work partially supported by FOM, The Netherlands},
        and 
        E.T. Tomboulis\address{Physics Department, Un. of California, Los 
                   Angeles, CA 90095-1547, USA}
             \thanks{Talk presented by E. T. Tomboulis at Lattice 2000, 
                    Bangalore, India. Work partially supported by 
NSF grant NSF-PHY 9819686.}}

\begin{document}

\begin{abstract}
We discuss the recently published numerical 
computations of the vortex free energy. They dramatically demonstrate that 
the excitation probability for a sufficiently thick vortex 
approaches unity at large beta, i.e. that thick vortices `condense' 
in the vacuum. This is known to imply confinement.  
An analytical approach for exploring this phenomenon 
is also outlined. 
\vspace{1pc}
\end{abstract}

\maketitle

\section{INTRODUCTION} 
A large body of recent work may be summarized 
as follows.  
Configurations  carrying nonzero vorticity over 
sufficiently large scales occur in the vacuum with nonzero 
probability for all $\beta < \infty$. 
In fact this probability approaches unity. The statement implies 
the presence of arbitrarily long thick vortices in the vacuum. 
This in turn automaticaly implies area-law for large Wilson loops,  
i.e. nonvanishing string tension for 
all $\beta < \infty$. 
 
A vortex carrying configuration may be characterized over a 
sufficiently large loop by a  
singular gauge transformation which is multivalued in $SU(N)$, 
but single-valued in $SU(N)/Z(N)$. The topological obstruction 
encountered in trying to extend the definition of the singular gauge 
transformation throughout the area enclosed by the loop forms then a 
surface of codimension 2. It is only the presence of the obstruction,  
rather than its  exact location (which is gauge dependent) that is 
of course significant, and signals the presence of a `vortex core'  
that cannot be deformed to the trivial vacuum by any continous deformation. 
It must be stressed, however, that in general such a configuration 
can have `core' vorticity distributed in a very nonlocal, patchwise manner. 
This in fact can be expected to be the case for 
generic individual configurations in the 
partition function sum. 
Individual vortex configuration with 
spacially well-defined localized core and long distance tail   
can of course be constructed and do occur  
in the path integral, but are not necessarily representative of most 
such configurations.  
One should not confuse individual vortex 
configurations in the partition sum with the canonical picture of the  
vortex realized as  a soliton of an effective long 
distance action (resulting from some integration over configurations).

This makes identification and computation of the contribution of 
vortex-carrying configurations in the path measure somewhat tricky. 
Thus, e.g., attempts to gauge fix so that vortices in 
any one configuration can be associated with well-localized thin 
`projection vortices' encounter ambiguities and problems that 
are discussed in other contributions in these proceedings \cite{lat00}.    
Here we report on 
recent studies \cite{KT1} which, for the first time, 
directly measured this contribution. 
The idea is to enforce the presence of a vortex (mod $N$) 
in every configuration  by an appropriate singular transformation,   
and measure the corresponding expectation. This then measures directly the 
contribution of such configurations, and hence the excitation 
probability (equivalently, the free energy cost) for a vortex in the vacuum.   
It provides the most direct and  
physically transparent method for assessing the   
presence of vortices in the vacuum at large $\beta$. 

\section{VORTEX FREE ENERGY} 
A coclosed set of plaquettes (2-cells) is a closed set 
of $(d-2)$-cells on the dual lattice. Thus, in $d=3$, 
it is a closed loop of dual bonds; in $d=4$, a closed 2-dim 
surface of dual plaquettes. Let 
$\V$ denote a coclosed set of plaquettes that 
winds through every, say, 2-dim $[12]$-plane of the lattice $\Lambda$, 
i.e. a topologically nontrivial plaquette set wrapped 
around the periodic lattice ($d$-torus) in the $(d-2)$ 
directions $\lambda \neq 1,\ 2$ perpendicular 
to $\mu=1\ \nu=2$. We define the partition function 
\bea
Z_\Lambda(\tau) &=& \int\prod_b dU_b\;\exp\bigg(
\,- \sum_{p\not\in \V} A_p(U_p) \nonumber\\
& & \qquad - \sum_{p\in \V} A_p(\tau U_p)\,\bigg) 
\label{tPF}\,,
\eea 
where the plaquette action $A_p(U_p)$ is replaced by the 
`twisted' action $A_p(\tau U_p)$ for each plaquette 
of $\V$. Here the `twist' $\tau \in Z(N)$ 
is an element of the center. There are thus $(N-1)$ different 
nontrivial choices for $\tau$. The trivial 
element $\tau =1$ is the ordinary partition 
function $Z_\Lambda(1) \equiv Z_\Lambda$. 
As indicated by the notation on the l.h.s.,  
the exact position or shape of $\V$ is irrelevant; 
the only dependence is on the presence of the $Z(N)$ flux 
winding through each $[12]$-plane. 
$\V$ can be moved 
around and distorted by a shift of integration variables, but 
not removed; it is rendered topologically stable by 
winding completely around the lattice. 
By the same token introducing two twists, $\tau$ on 
$\V$ and $\tau^{\,\prime}$ on $\V^{\,\prime}$,  
in (\ref{tPF}) is equivalent to introducing one twist 
$\tau^{\,\prime\prime}=\tau
\tau^{\,\prime}$ since $\V$ and 
$\V^{\,\prime}$ can be brought together by a shift of 
integration variables. This expresses the mod $N$ 
conservation of the $Z(N)$ flux introduced by the twist. Thus, 
for $N=2$, any odd number of such (nontrivial) twists is 
equivalent to one, and any even number to none.  

The twist enforces the introduction of 
a discontinuous (singular) $SU(N)$ gauge 
transformation {\it on every configuration} in the path integral 
measure in (\ref{tPF}) 
with multivaluedness in $Z(N)$; in other words, 
the introduction of a  
$\pi_1(SU(N)/Z(N))$ vortex. The set  
$\V$ represents the topological obstruction to 
having single-valuedness everywhere. 
Thus $Z_\Lambda(\tau)$, eq. (\ref{tPF}),  is the 
partition sum for configurations with a topologically stable 
vortex trapped inside and completely winding around 
the lattice. The vortex free energy (v.f.e.) order parameter, 
now defined as   
\bea
& & \exp(-F_{\rm v}(\tau))
 \equiv  {Z_\Lambda(\tau)\over Z_\Lambda} \nonumber\\
&  = &\vev{\exp\bigg(
- \sum_{p\in \V} \Big(A_p(\tau\,U_p)-A_p(U_p)
\Big)\bigg)}
,\label{mgfe}
\eea
is then the normalized expectation for the excitation 
of such a vortex. 

\begin{figure}[htb]
\begin{minipage}{75mm}
{\ }\hfill\epsfysize=6.5cm\epsfbox{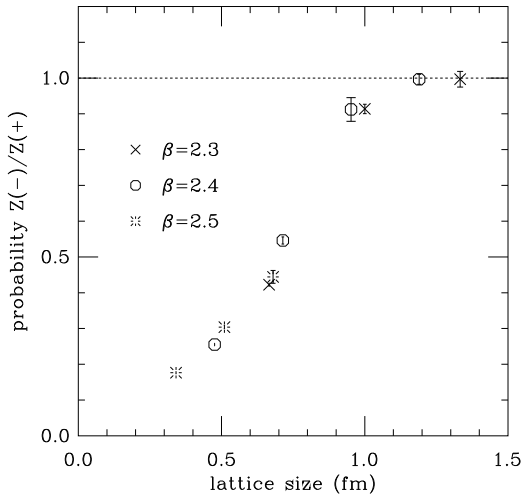}\hfill{\ }\\
\caption[tagv]{\label{vfe}$SU(2)$ vortex free energy 
(\ref{mgfe}) vs. lattice size }
\end{minipage}
\end{figure}

As it is evident from (\ref{mgfe}), measurement of this quantity 
presents a difficult sampling problem.  
It represents the expectation of an operator that 
can get significant contributions only 
from configurations having very small statistical weight, and 
thus becoming very difficult to measure with increasing lattice size. 
The problem can be addressed by use of a multihistogram method.   
Results of such a computation \cite{KT1} for $SU(2)$ are 
presented in figure \ref{vfe} for symmetric lattices and for 
three different values of $\beta$. The lattice spacings are 
$a=0.165$ fm, $a= 0.119$ fm  and $a=0.085$ fm for $\beta=2.3$, 
$\beta=2.4$, and $\beta=2.5$, resp. 
Notice that, with 
the lattice size expressed in physical units, the 
measurements for different $\beta$'s fall on the same curve, 
as they should. This indicates that the universal curve 
has been reached, and will not change at larger beta. 
Also, the onset of the sharp rise around $0.7$ fm is 
in the region of the finite temperature deconfining phase 
transition providing another indirect consistency check. 
The amplitude goes to unity for sufficiently large lattice size. 
For comparison, an upper bound for Coulomb massless behavior   
(from maximizing action in spin wave approximation) gives 
$\sim \exp (-\beta (\pi/2)^2) \approx 0.085$ at $\beta=2.3$.

Thus the  excitation free energy cost for a sufficiently thick 
vortex vortex goes to zero at large $\beta$. The curve gives a 
direct estimate of `sufficiently thick': slightly above $1$ fm. 
This is the thickness at which vortices can become arbitrarily 
long as the free-energy cost goes to zero.  
In this sense the vacuuum can be viewed as having a `condensate' 
of long thick vortices.  

Over distances well below $1$ fm the thick vortices ("confiners") 
are not readily detectable among perturbative and other shorter range 
fluctuations such as instantons dominating the action. (The instanton 
distribution is generally estimated to be centered around $0.3 - 0.4$ fm.)  
As a quantitative demonstration, let $<\tr U_p>^{(-)}$ denote the 
plaquette expectation in the presence of the twist, i.e. in 
the presence of a vortex winding around the lattice. For a   
$10^4$ lattice at $\beta=2.4$ we find  
$<\tr U_p>^{(-)} = 1.2557(2)$, while for the usual plaquette 
expectation in the absence of the twist   
$<\tr U_p> = 1.2598(2)$. Thus asymptotically in the large volume limit 
a vortex becomes locally `invisible'. 

\section{RELATION TO WILSON AND 'T HOOFT LOOPS} 

The behavior of other order parameters such as the Wilson and 
't Hooft loops can be related to that of the v.f.e. In particular, the 
existence of non-nonvanishing string tension is implied by 
confining behavior for the v.f.e. We consider $SU(2)$ for simplicity. 

Consider a rectangular loop $C$ in a $[12]$-plane.  
Then, for any reflection positive plaquette action, 
the Wilson $W[C]$ loop obeys the bound \cite{TY}:
\beq 
W[C] \leq \bigg(\,\exp(-\hat{F}) 
\,\bigg)^{\T{A_C\over A}} \;,  
\eeq 
where $A_C$ is the minimal area bounded by $C$, $A$ the area of 
the lattice $[12]$-plane, and 
\bea 
\exp(-\hat{F})& =& \sum_{\tau=1,-1}\; 
\;\tau\;\exp(-F_{\rm v}(\tau)) \nonumber\\ 
&=&\bigg(\,1 - \exp(-F_{\rm v}(-1))\,\bigg) 
\eea 
defines the $Z(2)$ Fourier transform  of the v.f.e. 
   
Similarly, let $B_{\rm min}$ denote a 't Hooft loop of minimal length, 
i.e. one consisting 
of the 4 cubes (in d=4) forming the 
coboundary of a plaquette (equivalently, the four dual bonds  
of the boundary of a dual plaquette), and having the flux 
attached to the loop winding completely around the lattice. 
This is nothing but (\ref{mgfe}) with one plaquette removed 
from the set $\V$ on which the $Z(2)$ twist resides. Physically, this 
may be viewed as an operator creating a vortex `punctured' by a $Z(2)$ 
monopole current loop of minimal length (a `tagged vortex'). 
Then it is not hard to prove that   
\beq 
B_{\rm min} \geq \exp(-F_{\rm v}(-1))\,\exp[-\beta\,<\tr U_p>^{(-)}] 
\label{lineq}
\eeq
where, as above, $<\tr U_p>^{(-)}$ denotes the plaquette expectation in 
the presence of the (unpunctured) twist winding around the lattice.  
Now confining behavior for $B_{\rm min}$ can be shown \cite{KT2} to 
provide a sufficient condition for the existence of a nonvanishing 
string tension through a lower bound on it. Hence, by (\ref{lineq}), 
the v.f.e. does also.

\section{AN ANALYTICAL APPROACH} 
It would clearly be very nice to get a (semi)analytic handle on the 
v.f.e. at large $\beta$. 
This is of course a very difficult problem. Here we would like to 
briefly outline a possible approach that we have been pursuing. 
Consider the standard polymer expansion of the partition function. 
It is based on the character expansion of the plaquette function 
Boltzmann factor:   
\[ e^{\frac{\beta}{2}\,\tr U_p}= c_0(\beta)\big[1 + \sum_i\,\frac{c_i(\beta)}{
c_0(\beta)}\,\chi_i(U_p)\big]. \] 
Inserting in $Z_\Lambda$ and expanding one obtains the polymer 
expansion: 
\bea
Z_\Lambda &=& \int \prod dU\;\prod_p e^{\frac{\beta}{2}\,\tr U_p} \nonumber\\
         &=& c_0(\beta)^{|\Lambda|}\Big[\,\sum_{\Gamma}\,\prod_\gamma 
z_\gamma\,\Big] 
\label{disc}\\
        &=& c_0(\beta)^{|\Lambda|} \exp\Big(\,\sum_C\,a(C)\,\prod_{\gamma\in C}
 z_\gamma\,\Big). \label{clus}
\eea 
The first sum (\ref{disc}) is over all sets $\Gamma$ of disjoint 
polymers $\gamma$ of activity $z_\gamma$; it is a {\it finite} series 
on a finite lattice $\Lambda$. The 
second sum (\ref{clus}) is over all `clusters' $C$, i.e. sets of 
overlapping polymers allowing multiple copies of the same polymer. 
There are simple explicit formulas for the coefficients $a(C)$.  
This is now an {\it infinite} series even on {\it finite} $\Lambda$.  
As it is well-known, the free-energy cluster expansion (\ref{clus}) 
converges for strong coupling (small enough $\beta$), and also in the 
$|\Lambda|\to\infty$ limit.

The v.f.e (\ref{mgfe}) now has a very nice  
expansion as the difference of two free energy cluster expansions  
\beq
 {Z_\Lambda(-)\over Z_\Lambda} = 
\exp\Big(\,\sum_C\,a_C\,\big(\prod_{\gamma\in C}
z_\gamma^{(-)} - \prod_{\gamma\in C}z_\gamma\big)\,\Big). \label{vclus}
\eeq 
Here $z^{(-)}$ are the activities in the presence of the twist. But 
(\ref{tPF}) is independent of the location at which the twist 
intersects any $[12]$-plane. Hence there is a huge cancelation of 
terms in the difference: only polymers that extend completely across 
$[12]$-planes contribute. But (\ref{vclus}) is, of course, still 
an infinite series even on a finite lattice.    

Proceeding in this manner at large $\beta$ appears hopeless 
unless one can work with a finite sum. Fornunately this can be 
accomplished by resuming all  
clusters with repeated multiplicities to 
convert the expansions (\ref{clus}), (\ref{vclus}) into a {\it finite} 
sum on a {\it finite} lattice. 
The result is   
\bea
\!\!\! & &\!\!\!{Z_\Lambda(-)\over Z_\Lambda}= \nonumber \\ 
\!\!\!\!& &\! \!\!\exp\Big(\,\sum_C\,\sum_{S\in S}\, a_{SC}\,
\big(\ln Z^{(-)}(S)  -  \ln Z(S)\big) \,\Big) \label{rvclus}
\eea  
where now the sum is over all skeleton clusters $C$ and all 
sub-skeleton clusters $S$ within each $C$, with $Z(S)$ the partition 
function for $S$, and explicitly computable coefficients $a_{SC}$.  
(\ref{rvclus}) is an exact expression that holds for all $\beta$. 

Having reached (\ref{rvclus}) as a suitable starting point, 
the next step is to define a decimation procedure for the sum over 
skeleton clusters on the lattice $\Lambda$ in terms of skeleton 
clusters on a coarser lattice $\Lambda^\prime$. Various decimation 
schemes for sets of clusters may be devised. If a sufficiently accurate 
such scheme can be implemented, it should, by successive iterations, 
reach the strong coupling regime without encountering a fixed point.    

\section{CONCLUSIONS} 

The question of confinement at  
large $\beta$ can be reduced to that of the excitation probability 
for an (arbitrarily long) vortex. Our recent numerical simulations have 
shown that this probability indeed tends to unity. 
Much work remains to be done. In particular, investigations on 
lattices of asymmetric length in the various directions are 
needed in order to obtain better understanding of the process 
of flux spreading leading to vortex thickening, as well as allow 
extraction of numerical values for the string tension directly from 
the v.f.e. Related work on 't Hooft loops, free energy derivatives and 
strong coupling effective actions is  
being pusued by several groups \cite{lat00}.

\end{document}